\documentclass[a4paper,fleqn,usenatbib,useAMS]{mnras}
\usepackage{graphicx}
\usepackage{natbib}
\usepackage{color}
\usepackage{amsmath}
\usepackage{amssymb}
\usepackage{multicol}   
\usepackage{bm}         
\usepackage{pdflscape}  
\usepackage{threeparttable}

\usepackage[dvipsnames]{xcolor}
\usepackage{color}

\newcommand{\ie}{i.\,e.}

\newcommand{\eg}{e.\,g.}

\newcommand{\kms}{\mbox{km\,s$^{-1}$}}

\newcommand{\nm}{\mbox{n\,m}}
\newcommand{\microm}{\mbox{$\mu$\,m}}

\newcommand{\teff}{$T_{\rm eff}$}

\newcommand{\fwhm} {{\sc FWHM}}

\newcommand{\EP} {\mbox{EP}}
\newcommand{\EW} {{\rm EW}}
\newcommand{\RV} {{\rm RV}}
\newcommand{\CCF} {{\rm CCF}}

\newcommand{\snr} {\mbox{SNR}}

\newcommand{\vmicro}{\mbox{$\xi_{\rm t}$}}
\newcommand{\gfeh}{\mbox{[{\rm Fe}/{\rm H}]}}
\newcommand\chem[1]{\mbox{\rm #1}}

\newcommand{\logg}{\mbox{log\,{\it g}}}
\newcommand{\loggf}{\mbox{log\,{\it gf}}}

\usepackage[T1]{fontenc}
\usepackage{ae,aecompl}

\title[Atmospheric Parameters from CCFs]{Atmospheric Stellar Parameters from Cross-Correlation Functions}

\author[L.\ Malavolta]{L.\ Malavolta$^{1,2}$\thanks{Corresponding authors: e-mail:
luca.malavolta@unipd.it (LM)},
C. Lovis$^{3}$,
F. Pepe$^{3}$,
C.\ Sneden$^4$,
S. Udry$^3$ \\
$^{1}$Dipartimento di Fisica e Astronomia ``Galileo Galilei'', Universit\`a di Padova, vicolo dell'Osservatorio 3, Padova IT-35122 \\
$^{2}$INAF - Osservatorio Astronomico di Padova, vicolo
dell'Osservatorio 5, Padova, IT-35122 \\
$^{3}$Observatoire de Genève, 51 Ch. des Maillettes, CH-1290 Sauverny, Switzerland \\
$^{4}$Department of Astronomy and McDonald Observatory,
The University of Texas, Austin, TX 78712, USA
}
\begin{document}
\pagerange{\pageref{firstpage}--\pageref{lastpage}} \pubyear{2017}

\maketitle
\label{firstpage}
\begin{abstract}
The increasing number of spectra gathered by spectroscopic sky surveys and transiting exoplanet follow-up has pushed the community to develop automated tools for atmospheric stellar parameters determination. Here we present a novel approach that allows the measurement of temperature (\teff ), metallicity (\gfeh ) and gravity (\logg ) within a few seconds and in a completely automated fashion.
Rather than performing comparisons with spectral libraries, our technique is based on the determination of several cross-correlation functions (CCFs) obtained by including spectral features with different sensitivity to the photospheric parameters. We use literature stellar parameters of high signal-to-noise (SNR), high-resolution HARPS spectra of FGK Main Sequence stars to calibrate \teff, \gfeh\ and \logg\ as a function of CCFs parameters.
Our technique is validated using low SNR spectra obtained with the same instrument. For FGK stars we achieve a precision of $\sigma_{\textrm{\teff}} = 50$ K, $\sigma_{\textrm{\logg}} = 0.09~ \textrm{dex}$ and $\sigma_{\textrm{\gfeh}} =0.035~ \textrm{dex}$ at $\textrm{\snr}=50 $, while the precision for observation with SNR $\gtrsim$ 100 and the overall accuracy are constrained by the literature values used to calibrate the CCFs.
Our approach can be easily extended to other instruments with similar spectral range and resolution, or to other spectral range and stars other than FGK dwarfs if a large sample of reference stars is available for the calibration. Additionally, we provide the mathematical formulation to convert synthetic equivalent widths to CCF parameters as an alternative to direct calibration.
We have made our tool publicly available.
\end{abstract}


\section{Introduction}

The advent of large-scale high-resolution spectroscopic surveys aimed at characterizing the properties of thousands to millions of stars, such as The Apache Point Observatory Galactic Evolution Experiment (\textit{APOGEE}, \citealt{Majewski2016}), the {\it Gaia}-ESO survey, and The GALactic Archaeology with HERMES (\textit{GALAH}, \citealt{Zucker2012}), has pushed the community towards the development of pipelines to automatically determine the fundamental parameters and chemical abundances of stars. Methods based on multivariate analyses relying on neural networks and machine learning are under constant improvement (see or example \citealt{Xiang2017} and references therein), alongside the more classical approaches of matching  the observed spectrum with a library of observed (\eg, \citealt{Yee2017}) or synthetic spectra (see \citealt{Endl2016} and \citealt{AllendePrieto2016} for a detailed review) or relying on the contrast or equivalent widths ratios of several spectral lines (\eg , see \citealt{Teixeira2016})

A widespread approach to measure the radial velocity (RV) of a star consists in cross-correlating the observed spectrum with a list of spectral lines in the wavelength rest-frame. It is reasonable to assume that the resulting cross-correlation function (CCF) reflects the mean shape of the observed spectral lines in the CCF list, which in turn depend on the photospheric parameters of the star. In fact, a similar path based on autocorrelation functions has been already followed in the past by \cite{Ratnatunga1989} and later by \cite{Beers1999} to estimate the metallicity of a large sample of stars. In this paper we demonstrate that it is possible to quickly and reliably measure  effective temperature \teff , gravity \logg , and metallicity \gfeh\ of a star by simply determining the CCF of its spectrum, when the chosen spectral lines are selected according to their sensitivity to the photospheric parameters.

The main advantage of the CCF approach is that no stellar continuum normalization of the spectrum under analysis is required, thus removing one of the most daunting and uncertain steps in photospheric parameter estimation (\eg, \citealt{Malavolta2014}). Our approach can deliver accurate and precise parameters for FGK main sequence stars within a few seconds and without any human intervention.

This paper is organized as follow. After a brief review of the so-called \textit{numerical} CCF technique, we derive an analytical formulation to link the CCF with the equivalent width (EW) of the individual lines used to build the CCF. We then introduce our method to calibrate the area of the CCF with respect to stellar atmospheric parameters from high-resolution and high \snr\ HARPS and HARPS-N spectra, obtained in both cases by  co-addition of several exposures at lower \snr\ with these instruments. Two different calibrations are presented. The first calibration is purely empirical and it provides effective temperature and metallicity from the observed CCF area for stars with parameters within the ranges $\textrm{\teff} \simeq [4500,6500]$ K, $\textrm{\logg} \simeq [4.2,4.8]$ and  $\textrm{\gfeh} \simeq [-1.0,0.5]$, \ie , solar-type stars. The second calibration is based on the transformation of synthetic equivalent widths of ionized elements into CCF areas and it allows the determination of gravity. Finally we validate the performance of our technique using the individual, low \snr\ exposures of each star obtained for exoplanet-search purposes, and determine the precision on the atmospheric parameters as a function of \snr .

\section{Cross-correlation function using a numerical mask}\label{sec:CCF_mask}

The CORAVEL-type cross-correlation function, or {\it numerical} CCF \citep{Baranne:1996tb,Pepe:2002ab}, consists in the numerical transposition of the optical cross-correlation made by instrument like CORAVEL \citep{Baranne:1979fm}, and it has been used in countless studies to measure the RVs of stars (\eg, \citealt{demedeiros2014} and references therein). The core of this technique resides in a binary mask where only the wavelength of the spectral lines to be included in the CCF computation have non-null values; the CCF is computed step by step for each point of the RV space by multiplying, in the wavelength space, the observed spectrum with the binary mask (shifted to the RV of the sampling point), and then integrating the result.  Following this definition, the outcome is a CCF where the minimum corresponds to the most likely RV of the star (\eg , \citealt{Mayor1995} and \citealt{Malavolta2015}).

A HARPS extracted spectrum is not usually ready for the CCF determination and some precautions must be taken in order to ensure that different exposures of a given star result in CCFs with the same characteristic parameters, after correcting for the observed radial velocity of the star. This is particularly true in our case, since we want to compare CCF parameters such as the Full Width at Half Maximum (\fwhm ) and the depth (also referred to as {\it contrast}) from different stars, and we must be sure that observed differences in the CCF parameters are intrinsic to the star and not due to the different observing conditions. The preparatory steps required before computing the CCF are described in the next section.

\section{Preparing the spectra for the CCF}\label{sec:correct_before_CCF}

The artifacts affecting a stellar spectrum can be divided in two main categories, \ie , the different observing conditions that affect the stellar light before entering the instrument, and the imprinting of the instrument itself on the observed spectrum. Instrumental effects are corrected by taking calibration images such as bias, dark and spectra of featureless sources, \eg , an halogen lamp. Nowadays modern data reduction pipelines perform this step automatically.

Changes in the observational conditions can affect the scientific outcome in many subtle ways. For example, due to differential refraction the position of the star in the focal plane of the telescope differs depending of its wavelength, with the effect getting worse at increasing airmass.
To compensate for this effect, modern telescope are provided with atmospheric dispersion correctors (ADC) which uses tabulated data to correct for the expected diffraction at telescope focus. Differences between tabulated and effective refraction at observing time  will still introduce a variation of the spectral distribution  (or simply spectral {\it slope}) of the stellar continuum. Additionally, the presence of aerosols or dust in the air, which absorption optical depth decreases with wavelength, can change the slope of the continuum by decreasing the \snr\ on the blue side of the spectrum.

Generally speaking, any effect that causes a variation of the spectral distribution will alter the CCF simply because less weight will be given to those lines that have suffered flux loss. Spectral reduction pipelines (such the Data Reduction Software of HARPS and HARPS-N) use a few standard templates to calibrate the flux distribution, but the choice of the template is usually left to the observer. When a star is observed to measure planet-induced RV variations, the only prerequisite is that all the observations should be corrected with the same arbitrary template. In this work we are looking for variations in the CCF as a function of the stellar parameters, so we must ensure that all the observations have been corrected for spectral distribution variations in an homogeneous way.

We decided to use a synthetic spectrum with Solar parameters as the reference template to determine the spectral slope function $a$ for a given observation (the detailed algorithm is described in Section~\ref{sec:refraction-corr}). This function describes the difference in spectral distribution of the star with respect to the reference template, and at the same time it compensate for variations in the spectral slope due to changing observational conditions.
We denote with $f_s$ a stellar spectrum $f$ that has been corrected for the spectral slope, and with $c$ and $c_s$ the two functions required to normalize  $f$ and $f_s$ respectively to unitary flux over 1 \AA , as commonly done before performing equivalent width measurements.

After spectral slope correction, the continuum function $c_s$ of the science spectrum is related to the template flux-calibrated continuum $c_t$ through a scaling factor $m$, which is simply the ratio between the template flux and the corrected science flux at the reference wavelength $\lambda _{\rm ref}$ (Equation \ref{eq:m_definition}).

\begin{equation}\label{eq:m_definition}
\begin{split}
& c_s(\lambda)  = m c_t(\lambda) \\
& m = \frac{f_s(\lambda_{\rm ref})}{f_t(\lambda_{\rm ref})} =  \frac{a(\lambda) * f(\lambda_{\rm ref})}{f_t(\lambda_{\rm ref})}
\end{split}
\end{equation}

As a consequence, the stellar continuum function is given by Equation \eqref{eq:continuum_def}.
\begin{equation}\label{eq:continuum_def}
\begin{split}
c(\lambda,p) & = m c_t(\lambda) /  a(\lambda) \\
\end{split}
\end{equation}

The advantage of using the factor $m$ and Equation \eqref{eq:continuum_def} is that now there is no need to determine the local continuum of the observed spectrum in order to compute the CCF from spectra obtained with varying observing conditions. Since the continuum normalization is the most difficult task in the analysis of low \snr\ spectrum, the advantage is considerable. The drawback is that now the obtained values will depend on the goodness of the spectral slope correction in restoring the correct flux as a function of wavelength. This task however can be easily performed, as we will show in Section \ref{sec:refraction-corr}.

\section{Correction of spectral distribution variations}\label{sec:refraction-corr}
\begin{figure}
\includegraphics[width=\columnwidth]{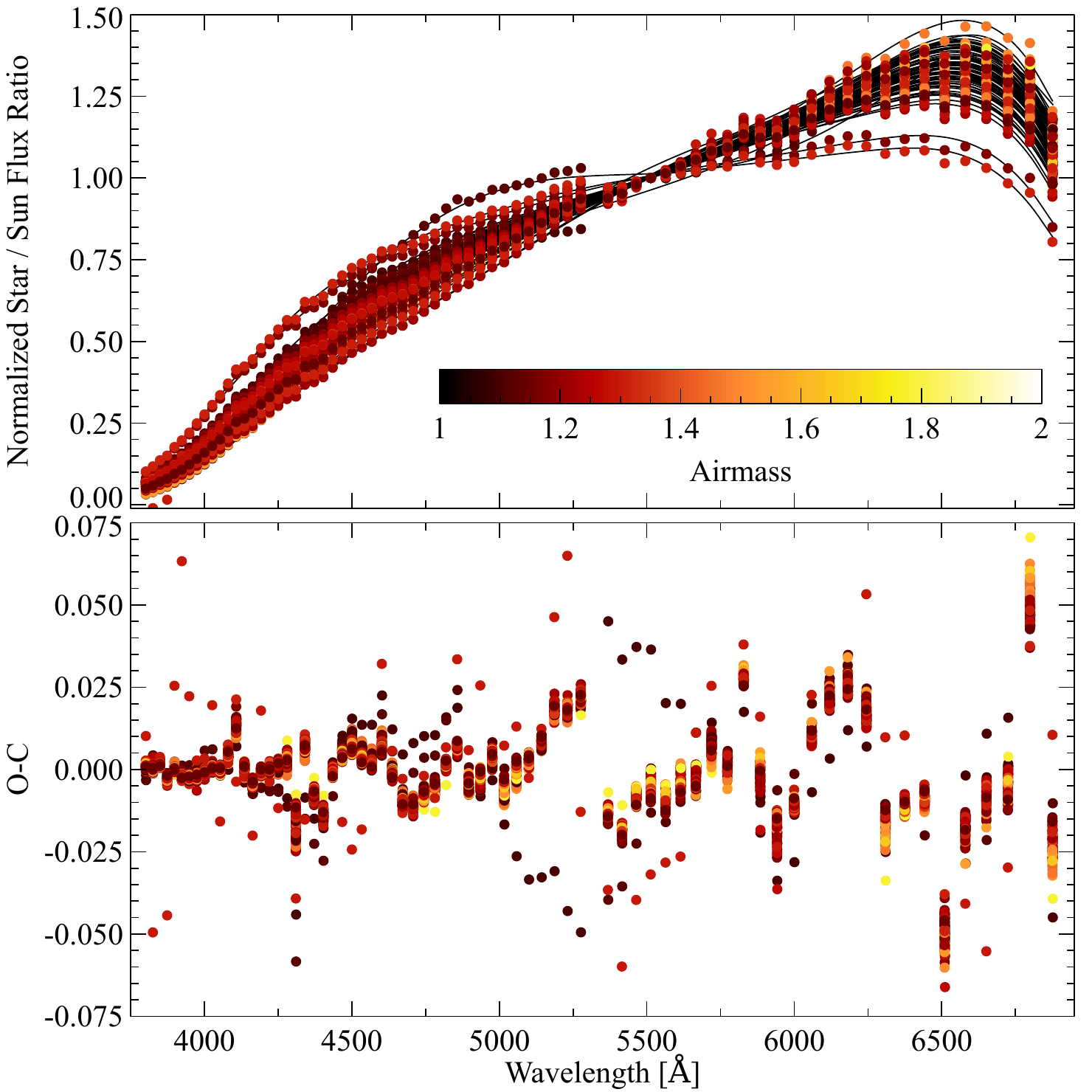}
\caption{Upper panel: Normalized flux ratio with respect to a synthetic template for 75 exposures of HD125612. Each point represents the value computed form an individual order. Points are color-coded according to the airmass of each exposure. The continuous line represents the correction factor as a function of wavelength for each exposure. Bottom panel: residuals between the star-template ratios and the corresponding correction function. }
\label{fig:fluxcorr_star}
\end{figure}

Here we describe a general algorithm to bring the observed spectra to a standard reference flux distribution, in order to remove differences between exposures due to changes in observing conditions, as introduced in Section~\ref{sec:correct_before_CCF}.
The correction is performed using a spectrum with $\textrm{ \teff } = 5750 $ K, $\textrm{\logg} = 4.5$ and $\textrm{\gfeh } = 0.00$ from  the synthetic stellar library of \cite{Coelho:2005ab}. This library provides the continuum stellar flux for each spectrum; it has high resolution spectra ($R>200000$) and it
covers a wide range of wavelengths (from $300$ \nm\ to $1.0$ \microm ). Any other library with similar characteristics works equally well.
The algorithm to obtain the flux correction function $a(\lambda)$ for an echelle spectrum includes the following steps:
\begin{itemize}
  \item Before proceeding, the spectra must be corrected for the blaze function, to remove order-dependent instrumental effects.
  \item For each extracted order the average flux for unit wavelength is determined. Integration is performed on the central half of the extracted order (from \textrm{pixel} 1024 to 3072 for HARPS spectra); the mean wavelength between the two integration limits is taken as reference value for the wavelength.
   \item The same operation is performed on the template spectrum.
   \item The factor $m$ (Equation \ref{eq:m_definition}) is obtained by dividing the integrated flux from the reference order (in our case, the order centered on $5500$ \AA ) of the observed spectra by the respective value of the template.
   \item The observed-derived values are divided by $m$ and by the respective template-derived values to determine the normalized star-sun ratio for each wavelength step.
   \item The correction factor as a function of wavelength is obtained by interpolating the normalized star-template ratios with a B-spline of the 4th order.
 \end{itemize}
We noticed that some points in our HARPS spectra have a correction factor systematically lower in some wavelength domains, \eg , in the 3900-4000 \AA\  region: this is due to the presence  of strong lines (in the mentioned region, the CaII H-K doublet) where the differences between the observed spectra and the template become important; the affected orders are excluded from the fit in the last step.
An example of the procedure is shown in Figure~\ref{fig:fluxcorr_star}, where the flux correction has been applied to 75 exposures of the star HD125612.

The resulting function is a combination of the systematic spectral difference between the star and the template, a general trend caused by the wavelength-dependent light loss due to either the telescope (\eg , mirror coating) or the instrument (\eg , CCD sensitivity curve), and small variations resulting from the difference between actual atmospheric refraction effect and the one predicted by the ADC for a given airmass.

\section{The link between the CCF and the Equivalent Width}\label{sec:link_CCF_EW}

Determination of atmospheric parameters is based on the measurement of the equivalent width (\EW ) of spectral lines, and the numerical CCF is the direct transposition of spectral lines from wavelength to radial velocity space. We need to find an analytical formulation to compute the area subtended by the CCF starting from a set of \EW s.

Spectral lines on the linear part of the curve of growth are usually approximated with a gaussian shape \citep{Gray:2005bb}. An \EW\ is obtained by measuring the depth of a line in the normalized spectrum $d$ (also known as the {\it contrast} of the line) and its full width half maximum $\Gamma$ in the wavelength space. Radial velocities of stars are always well below the relativistic limits (\ie , $\RV_{star} <1000 $ \kms ), so we can use the classical Doppler formula to transpose the \EW\ in the radial velocity space, as in Equation \eqref{eq:area_ew1}.

\begin{equation}\label{eq:area_ew1}
\begin{split}
  \EW & =  \frac{ \sqrt{2 \pi} }{2 \sqrt{2 \ln{2}}  }  d_{\lambda} \Gamma_{\lambda} \\
  & = \frac{ \sqrt{2 \pi} }{2 \sqrt{2 \ln{2}}  } \frac{\lambda_0}{c} d_{\RV} \Gamma_{\RV} \\
  & =  \frac{\lambda_0}{c} A_{\CCF }
\end{split}
\end{equation}

The normalized area $A_{\CCF}$ subtended by the CCF is equivalent to the \EW\ of Equation \eqref{eq:area_ew1} only when a single line is considered. Following \cite{Pepe:2002ab}, the observed CCF is obtained by coadding the individual CCFs from each line in the mask, without a prior normalization for their respective continuum. Consequently the corresponding area $A_{\CCF}$ is given by the sum of the areas of each spectral line $i$ in the CCF mask weighted by their continuum levels in the RV space.
The value of the continuum level at the center of the CCF (\ie , $\RV = 0 $) for a generic line $i$ is obtained by applying the definition of the CCF to the continuum function $c(\lambda )$ at the wavelength $\lambda _i$ corresponding to the center of the spectral line, with $\delta^i_\lambda$ being the size of the bin used to compute the CCF, as in Equation \eqref{eq:ccf2ew_step1}\footnote{The integration bin is usually taken  constant in the RV space, so in the wavelength space its value depends on the line under analysis}.
\begin{equation}\label{eq:ccf2ew_step1}
c^i_{\CCF }(0) =  \int_{\lambda_i - \delta
  ^i_\lambda/2}^{\lambda_i +
\delta ^i_\lambda/2} c(\lambda) d\lambda \simeq \delta _\lambda c(\lambda_i)
\end{equation}
In a generic case it would be difficult to obtain the integrand from Equation \eqref{eq:ccf2ew_step1}. In practice, $\delta _\lambda$ is always chosen to be very small compared to the instrumental FWHM, so we can approximate $c(\lambda)$ as constant between the integration limits.

Using Equation \eqref{eq:continuum_def} from Section \ref{sec:correct_before_CCF}, the expected CCF area as a function of the \EW s of the spectral lines in the CCF mask can be determined (Equation \ref{eq:ccf2ew_final}).

\begin{equation}\label{eq:ccf2ew_final}
\begin{split}
  A_{\CCF} & = \frac{\sum_i \delta^i_\lambda c(\lambda_i)  A_i}{c_{\CCF }(0)} \\
  & = \frac{1}{c_{\CCF }(0)} \sum_i {\frac{  \delta^i_\lambda m c_t(\lambda_i)}{a(\lambda_i)} A_i } \\
   & = \frac{1}{c_{\CCF }(0)} \sum_i {\frac{  \delta^i_\lambda m c_t(\lambda_i)}{a(\lambda_i)} \frac{\lambda_i}{c} \EW _i }
\end{split}
\end{equation}

All the variables in the right side of Equation \eqref{eq:ccf2ew_final} are known a priori  ($c_t$, $\delta^i_\lambda$) or can be obtained by the observations themselves ($m$, $a$, $c_{\CCF }(0)$ using the sides of the CCF function), while the \EW\ for each line $i$ in the CCF mask can be computed  with a spectrum synthesis code. This result allows the determination of the expected CCF area as a function of the atmospheric parameters and the creation of a synthetic calibration.

\section{Characteristic of the sample of stars}\label{sec:sample_stars}

{\ bf It is essential to have a calibration sample of stars that widely span the range in the stellar parameters that we want to calibrate.} At our disposal we have $1111$ stars from several HARPS long-term programs \citep{Mayor:2003wv,LoCurto:2010aa,Santos:2011ds}.
Since these stars are the targets of extensive exoplanet search surveys, a large number of exposures have been collected during several years of HARPS operations: this allows us to use the same stars as
\textit{calibrators} when considering the coadded, high \snr\ spectrum, and as \textit{test case} when a single low \snr\ spectrum is under study.

\begin{figure}
\includegraphics[width=\columnwidth]{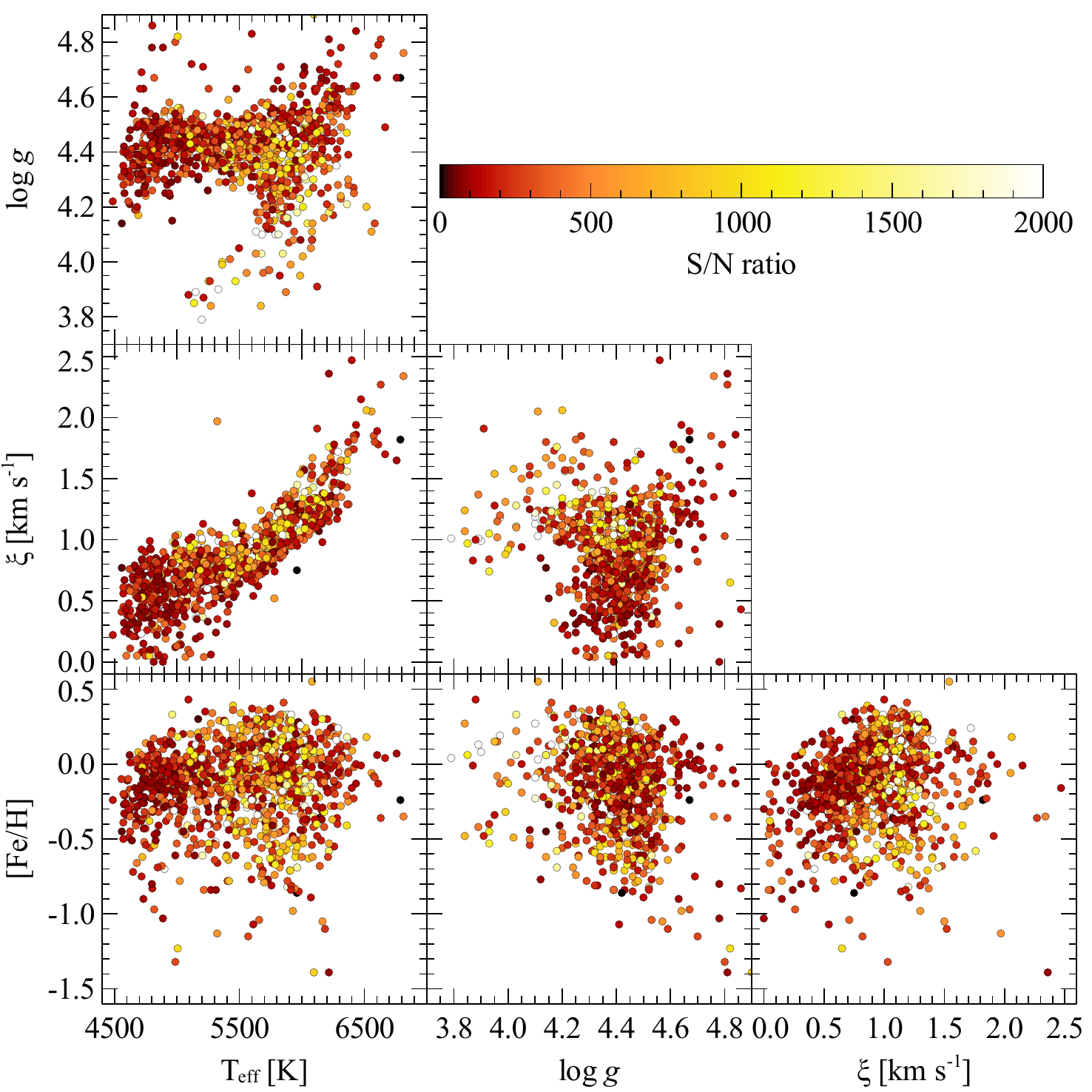}
\caption{The stellar atmosphere parameters of the \textit{calibration}  sample, derived by \citet{Adibekyan:2012ab} using high \snr\ spectra. Except for \vmicro , no trend between atmosphere parameters is present.}
\label{fig:star_sample}
\end{figure}

Stellar atmosphere parameters have been derived by \cite{Adibekyan:2012ab} from co-added spectra using classical \EW\ analysis. They employed empirical transition probabilities using Solar spectra as reference, thus making their analysis differential with respect to the Sun. The automatic continuum determination and \EW\  measurements performed with \texttt{ARES} \citep{Sousa:2007gn}, ensure a high degree of homogeneity in the parameters derived for this sample. The distribution of the atmosphere parameters are displayed in Figure~\ref{fig:star_sample}: the sample contains only dwarf stars with $\overline{\logg} \simeq 4.4$, with a few exceptions.
The targets span a wide range in \teff\  and \gfeh\ without any correlation between the two parameters.
A clear trend of micro-turbulence \vmicro\ with effective temperature is clearly visible. Since this relationship is well known and has been studied in the past (see for example the calibration of \citealt{Tsantaki2013}) we will not use the CCFs to derive this parameter. The quoted errors are $\sigma_{\textrm{\teff}} = 30~$ K, $\sigma_{\textrm{\logg}} = 0.06~ \textrm{dex}$ and $\sigma_{\textrm{\gfeh}} = 0.03~ \textrm{dex}$ on average. These uncertainties are internal of the method and do not include systematic source of errors such as the use of a specific set of stellar atmosphere models or departure from local thermodynamic equilibrium (LTE).

\section{Line selection for the CCF masks}\label{sec:line_selection_CCF_mask}

The HARPS Data Reduction Software (DRS) automatically determines an overall CCF and its parameters (central RV, \fwhm\ and contrast) at the end of each exposure using one of the CCF masks provided by the DRS. Each mask contains several thousands of lines chosen among several chemical elements and ionization states and weighted according to the amount of radial velocity information they carry. For example, deep and sharp lines have larger weights simply because their RVs are easier to measure \citep{Pepe:2002ab}. Three CCF masks are available, based on synthetic spectra of G2, K5 and M2 dwarf stars, in order to take into account the variation of line depths and flux distribution with the spectral type of the star\footnote{Detailed information regarding the HARPS DRS can be found at the instrument website \url{http://www.eso.org/sci/facilities/lasilla/instruments/harps.html}}.
Since every line carries information on the radial velocity shift of the star regardless its chemical origin, and the principal goal of the instrument is to measure the differential shift of the star to a precision better than $1 \textrm{m}/\textrm{s}$, all the available unsaturated lines are included in each mask in order to increase the \snr\ of the CCF.

In this work we are pursuing a different goal: to determine a correspondence between the characteristics of the CCF and the atmosphere parameters of the stars.
For our purposes, significant noise can be added to the derived relationships by inclusion of lines with unknown atomic parameters or from  chemical elements that can have large star-to-star abundance variations.
We must make a more stringent selection of the lines included in the CCF mask in order to preserve the relationship between photospheric parameters and CCF characteristics without sacrificing excessively the \snr\ of the CCF.

To select a good sample of lines for our mask, we started with the list from \cite{Malavolta2016}. This list includes 498 atomic line parameters from \cite{Sousa:2011gr}, which is in turn an extension of the iron line list from \cite{Sousa:2007gn}, and several chemical elements from \cite{Neves:2009jz}. Transition probabilities \loggf\ have been updated to match the Solar \EW s with the elemental abundances of \cite{Asplund2009}. Our results will be then differential with respect to the Sun.

We briefly describe the procedure followed to derive the line list. The initial line list, along with the atomic parameters (including an initial estimate of the oscillator strengths) is taken from the
\textrm{VALD4} online database (\citealt{Piskunov:1995ui}, \citealt{Kupka:2000aa})
 using the solar stellar parameters as input ($\textrm{\teff}=5777$ K,
 $\textrm{\logg}=4.44$, $\xi =1.0 $ \kms ) in the spectral region from 4500 \AA\, to 6910 \AA . The following criteria are followed:
 \begin{itemize}
 \item Lines must not be strongly blended in the Kurucz Solar Flux Atlas (\citealt{Kurucz05} and references therein).
 \item Lines must not be too weak ($\textrm{EW}<5 \,m$\AA ) or too strong
   ($\textrm{EW}>200 \,m$\AA)
 \item Lines must not reside in the wings of strong lines (\eg H$\alpha$, H$\beta$ and MgI lines)
 \item \EW s measured in the  Kurucz Solar Flux Atlas and in the solar reflected light spectrum of the Ceres asteroid (obtained with HARPS) must agree within 10\%.
 \item \EW s of the lines can be appropriately measured using the  \texttt{ARES} program \citep{Sousa:2007gn}.

 \end{itemize}
Finally empirical oscillator strengths are obtained through an inverse analysis with the \texttt{ewfind} driver of the LTE Spectral Synthesis code
\texttt{MOOG}  \citep{Sneden:1973el} and the measured \EW s from the solar spectrum, assuming for the Sun the parameters listed in \cite{Santos:2004gr}.

The provided wavelengths are used as input positions by the \texttt{ARES} program to perform a Gaussian fit of the spectral lines in order to derive their \EW s. However these wavelengths lack the required precision to be directly used in a cross-correlation mask, \ie , spectral lines are provided with a precision of $0.01$ \AA , corresponding to a potential misplacement of the CCF of $0.545$ \kms\ at $5500$ \AA . We re-determined the central wavelengths of the lines by performing a multi-gaussian fit of each spectral line in the list on several solar spectra observed with HARPS\footnote{Available at \url{http://www.eso.org/sci/facilities/lasilla/instruments/harps/inst/monitoring/sun.html}}. The new values are already included in Table~\ref{tab:sousa2010_linelist}.

\begin{figure}
 \includegraphics[width=\columnwidth]{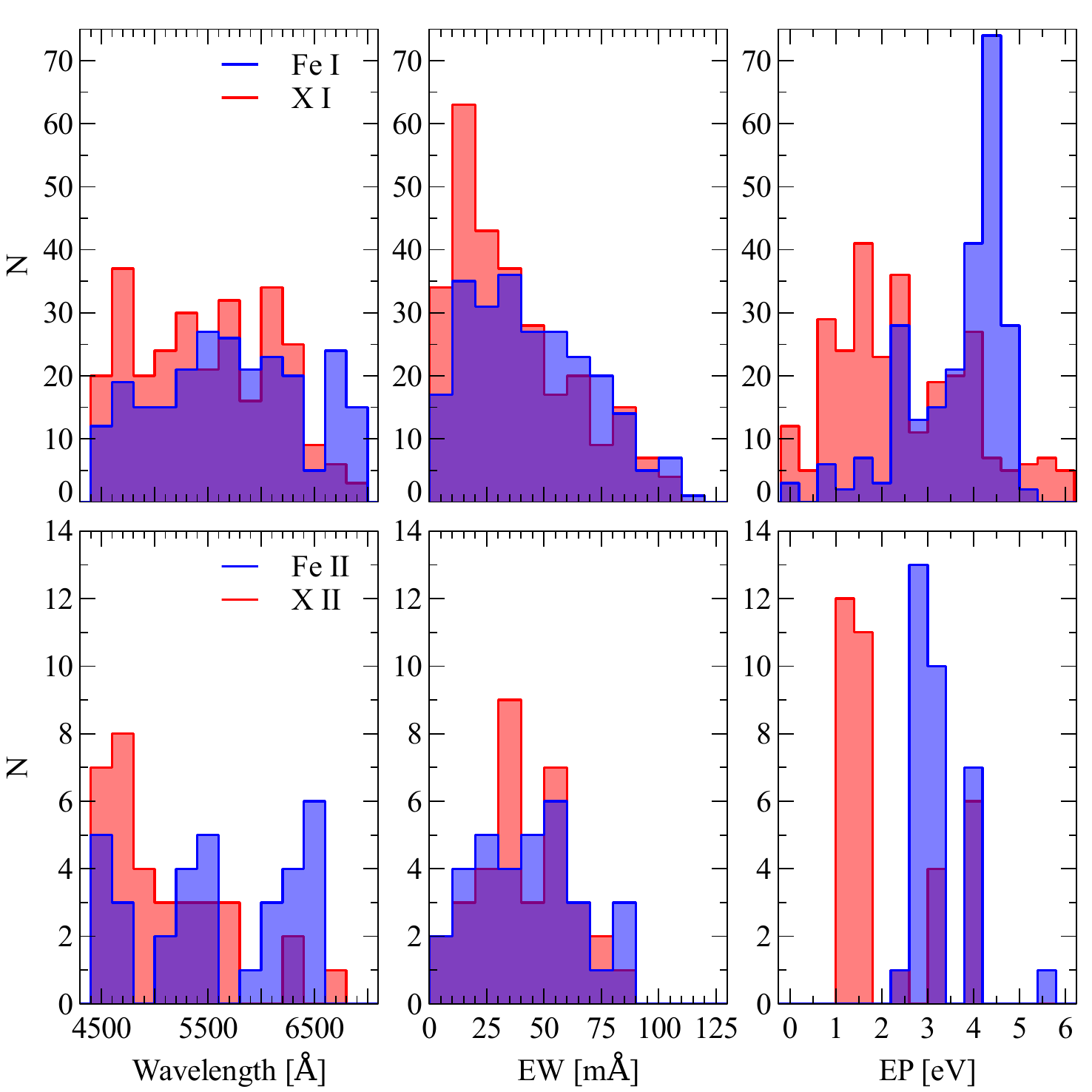}
\caption{Distribution of selected neutral lines (upper panels) and ionized lines (lower panels) of Fe (blue histograms) and other ``X'' elements according to their wavelength, equivalent width (EW) measured in the solar spectrum and excitation potential (EP). The blue and red histograms are plotted separately, leading to overlapping (purple) parts of the histograms. }
\label{fig:linelist}
\end{figure}

In Figure~\ref{fig:linelist} the distributions of the selected lines versus the basic atomic line parameters are displayed. The selected lines are homogeneously distributed in wavelength and cover a good range in excitation potential (EP) and equivalent width as measured in the solar spectrum.

\begin{figure}[h]
 \includegraphics[width=\columnwidth]{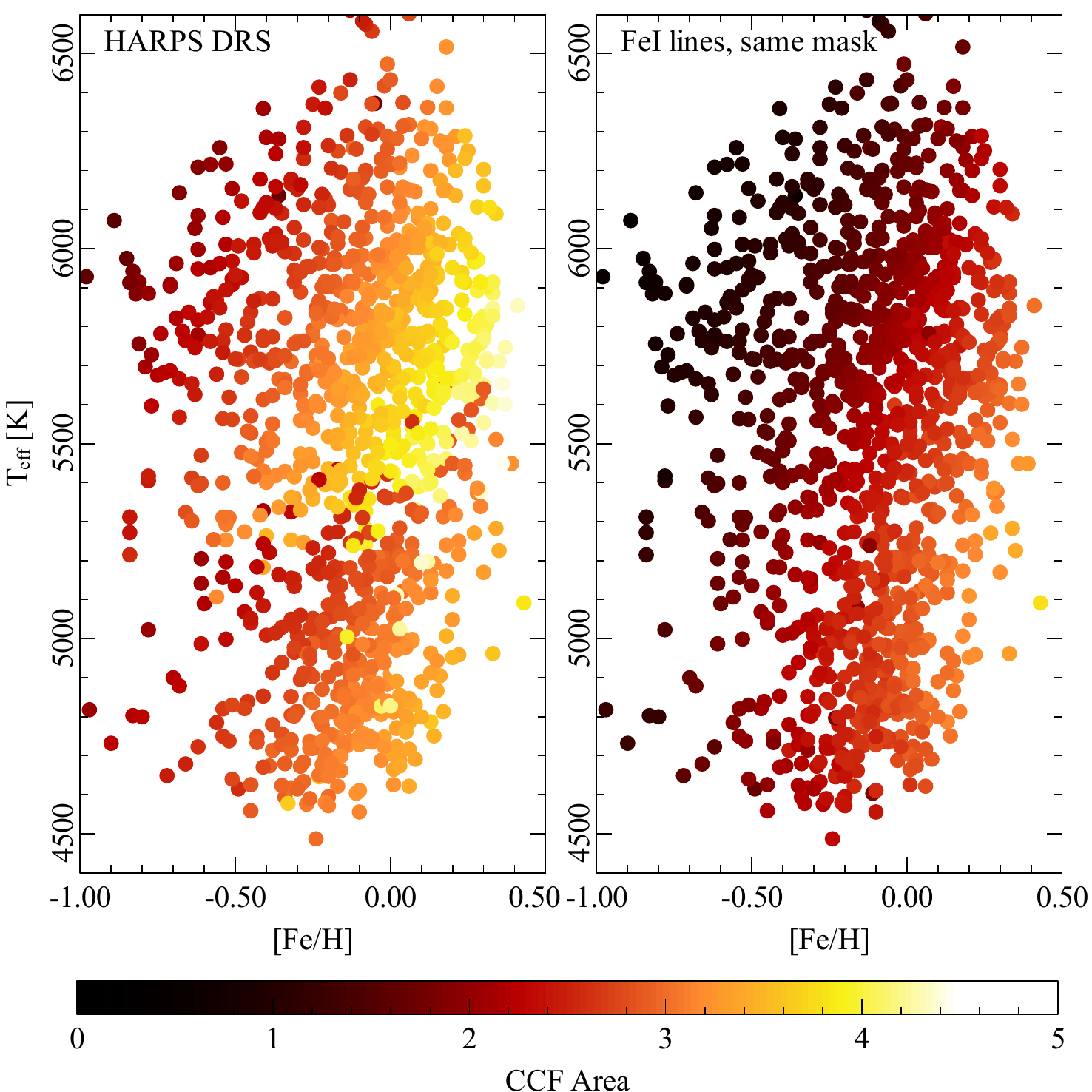}
\caption{In the left panel, the CCF Area in function as determined by the HARPS DRS in the \teff - \gfeh\ plane; the jump in the area values is due to the different mask and spectral template used by the DRS for a given star. In the right panel, the area of the CCF as a function of the atmospheric parameters when using only \chem{FeI}  lines. In both panel, it is evident a degeneracy between \teff\ and \gfeh\ for a given CCF area.}
\label{fig:compareCCF}
\end{figure}

In Figure~\ref{fig:compareCCF} the measured continuum-normalized area of the CCF determined by the HARPS pipeline (left panel) as a function of
effective temperature and metallicity is compared with the CCF area obtained when only \chem{FeI} lines are
used (right panel), for the same set of stars in Figure \ref{fig:star_sample}.
The jump in color in the left panel is due to the mask and flux correction template used for different spectral types, and it is the main reason why we cannot use the CCF information provided by the DRS pipeline. The area of the CCF varies smoothly when it is homogeneously derived for all the star in the sample.
In both panels a degeneracy between temperature and metallicity is visible for a given CCF area. Every line depends on both the effective temperature of the photosphere and the abundance of the chemical element that is producing the line, and this behavior is conserved when the lines are co-added into a CCF. It is clear then that a single CCF area is not enough to constrain both metallicity and temperature if available.
This is true regardless of the binary mask used for the CCF construction.

Fortunately, spectral lines react to temperature changes in different ways according to their atomic parameters.
In classical stellar atmospheric parameters determination, a key parameter when determining the effective temperature of the star is the excitation potential (\EP ), \ie , the energy required to excite an atom to a given state from the ground state.

\begin{figure}
 \includegraphics[width=\columnwidth]{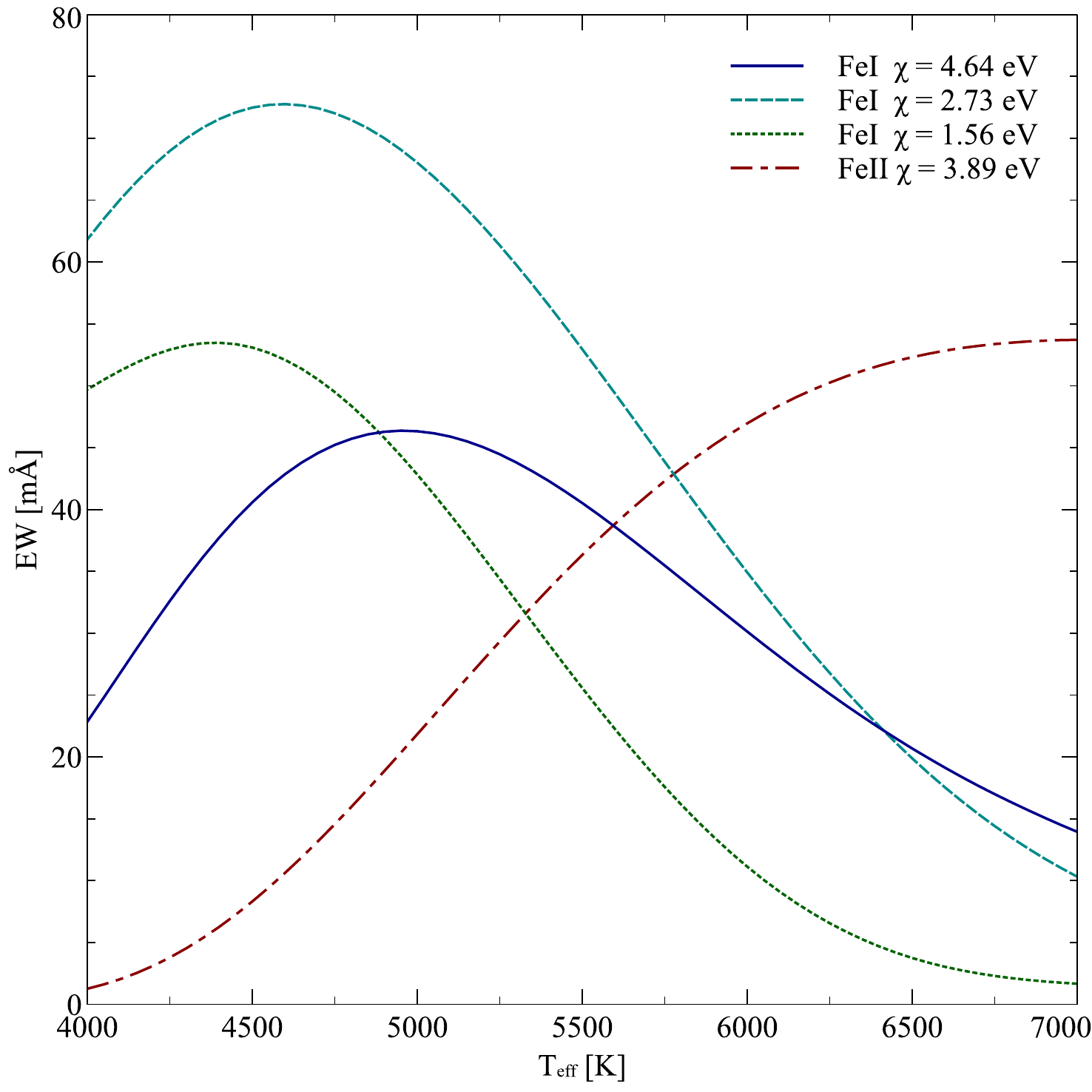}
 \caption{Theoretical \EW s curve for Iron lines with different  excitation potentials and ionization state as a function of the effective temperature. Values have been calculated using Kurucz atmosphere models with \logg , \vmicro , and \gfeh\ fixed at the Solar values.}
 \label{fig:Teff_EP_FeI}
 \end{figure}

The presence of a trend between the abundances of individual Fe~{\sc I} lines (directly derived from their \EW s) and the excitation potential is a clear sign of an incorrect \teff\ assumed for the stellar model.
The relationship between temperature and theoretical equivalent widths for lines with different excitation potentials is demonstrated with several examples in Figure~\ref{fig:Teff_EP_FeI}. For each value of the effective temperature the \EW s have been determined using a Kurucz model atmosphere with $\logg=4.44$, $\vmicro=1.0$ \kms\ and $\gfeh=0.00$ using the
 \texttt{ewfind} driver of \texttt{MOOG}. Calculations have been performed for three neutral Fe lines and a ionized one. Helpful analytical expressions for \EW\ curves as functions of temperature, excitation and ionization potentials are derived in \cite{Gray:2005bb}.

 We can take advantage of the fact that at a given temperature the \EW\  curve of each line has a different slope, depending on its excitation potential, to create CCF masks whose associated areas have different gradients in the temperature-metallicity plane of Figure~\ref{fig:compareCCF}. We split our original linelist into three CCF masks:

(i) a mask with low excitation potential lines, comprising 243 lines
  with $\EP<3.0$ eV ;
(ii) a mask with 94 intermediate \EP\ lines ($3.0 < \EP < 4.0$ eV);
(iii) a mask with 183 high
excitation potential lines ($\EP>4.0$ eV)

The list of spectral lines used in this work, their atomic parameters, and their mask membership are given in Table~\ref{tab:sousa2010_linelist}. The values obtained with these three mask will be identified respectively by the symbols $A_{low}$, $A_{med}$ and $A_{high}$ .

\begin{table*}
\center
\caption{Chemical elements, atomic parameters, solar \EW\ from \citet{Adibekyan:2012ab}, contrast value of the spectral line as observed in the Solar spectra, and mask membership for all the spectral lines used in this work. Only a sample is given here, the full table will be available in the online version of the paper.}
\label{tab:sousa2010_linelist}
\begin{threeparttable}
\begin{tabular}{ccccccc}
 \hline
 Chemical species  &  Wavelength [\AA ] & \EP & \loggf & $EW_{sun}$ & Contrast & Mask\tnote{a} \\
\hline
 FeI &    4554.462  &  2.870  &  -2.752  &   42.6  &  0.350 & L \\
 FeI &    4561.413  &  2.760  &  -2.879  &   40.0  &  0.368 & L \\
 FeI &    4574.722  &  2.280  &  -2.823  &   63.8  &  0.621 & L \\
... & ... & ... & ... & ... & ... & ...  \\
\hline
\end{tabular}
\begin{tablenotes}
  \small
  \item[a] L: low \EP\ lines; M: intermediate \EP\ lines; H: high \EP\ lines; I: ionized lines.
 \end{tablenotes}
\end{threeparttable}
\end{table*}

\begin{figure}
\includegraphics[width=\columnwidth]{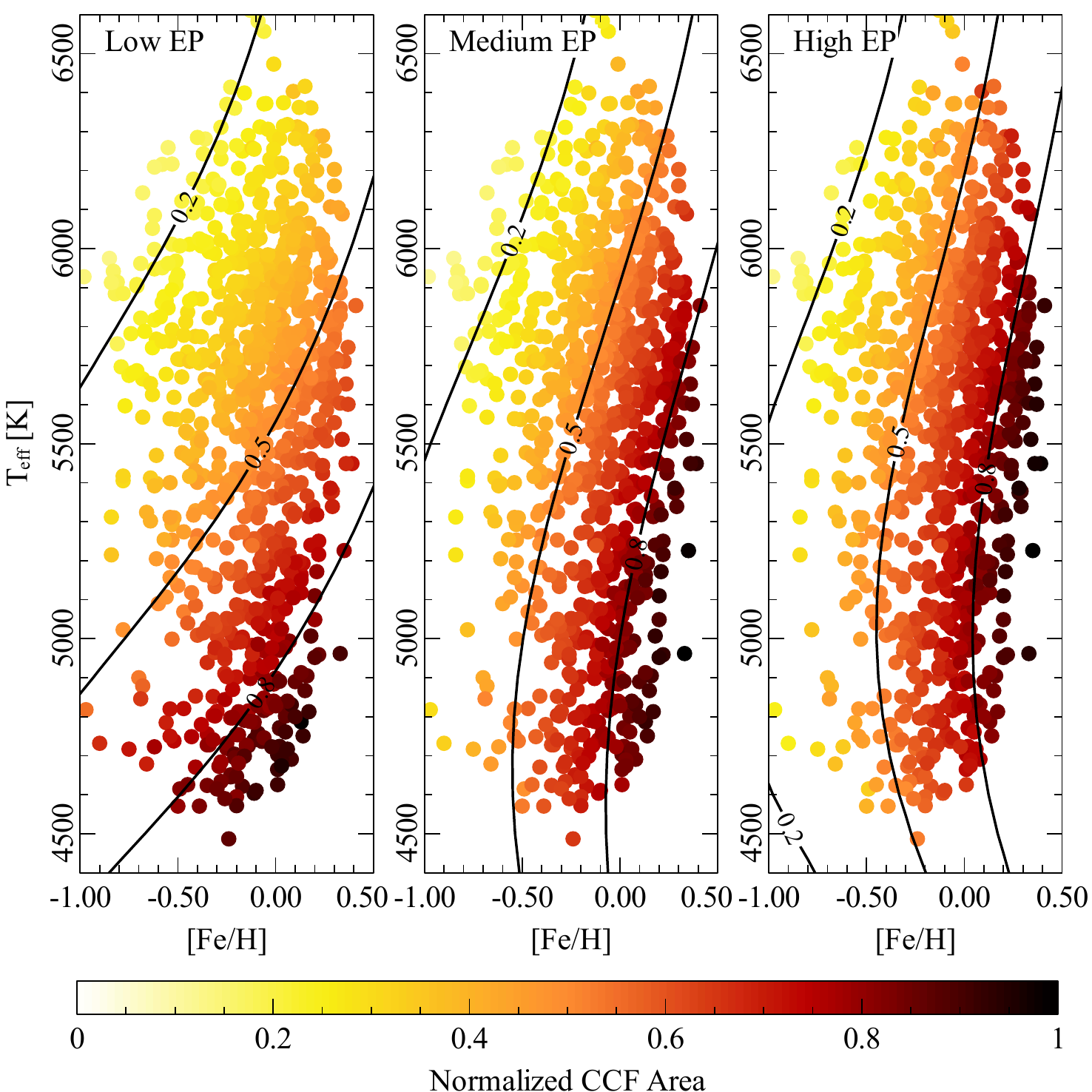}
\caption{CCF areas as a function of temperature and metallicity, using the three masks defined in Section~\ref{sec:line_selection_CCF_mask}. Areas have been normalized to the maximum value of each mask for illustrative purpose only. Iso-area lines with values 0.2, 0.5 and 0.8 have been drawn to highlight the behaviour of the CCF areas derived with different masks.}
\label{fig:tg_plane_area}
\end{figure}

In Figure~\ref{fig:tg_plane_area} we can see that the iso-areas lines have different slopes for each mask, with stronger slope variations for cooler stars. As a result, a given combination of CCF areas will identify an unique point in the temperature-metallicity plane, \ie only a single $($\teff,\gfeh$)$ pair can  match the observed CCF Area, thus breaking the \teff - \gfeh\ degeneracy.

 \section{Modeling the Temperature-Metallicity plane}\label{sec:model-temp-gfeh}

In the temperature range of FGK dwarf stars most of the elements are in their first ionization stage, so weak lines formed by neutral elements are insensitive to pressure changes \citep{Gray:2005bb}. In the previous section we described CCF masks composed of neutral-species transitions, with which we can derive an empirical calibration for \teff\ and \gfeh\ which is suitable for dwarf stars (\logg\ $\simeq 4.4 \pm 0.3 \,\textrm{dex}$) without requiring a precise knowledge of stellar gravity.

The easiest way to determine \teff\ and \gfeh\ from the available CCF areas is to calibrate the two parameters as functions of the available CCF Area from high \snr\ spectra:
\begin{equation}
\textrm{\teff}   =
f_1(A_{\textrm{low}},A_{\textrm{med}},A_{\textrm{high}})
\label{eq:teff_func1}
\end{equation}

\begin{equation}
\textrm{\gfeh} =
f_2(A_{\textrm{low}},A_{\textrm{med}},A_{\textrm{high}})
\label{eq:gfeh_func1}
\end{equation}

The two functions $f_1$ and $f_2$ can be represented in any form, \ie , either an analytical function or a list of tabulated values to be interpolated.

The coefficients of the functions are determined through Least Squares minimization using as input data  \teff\ and \gfeh\ from literature, and the CCF areas of the three masks measured on high \snr\ spectra.

\begin{figure}
\includegraphics[width=\columnwidth]{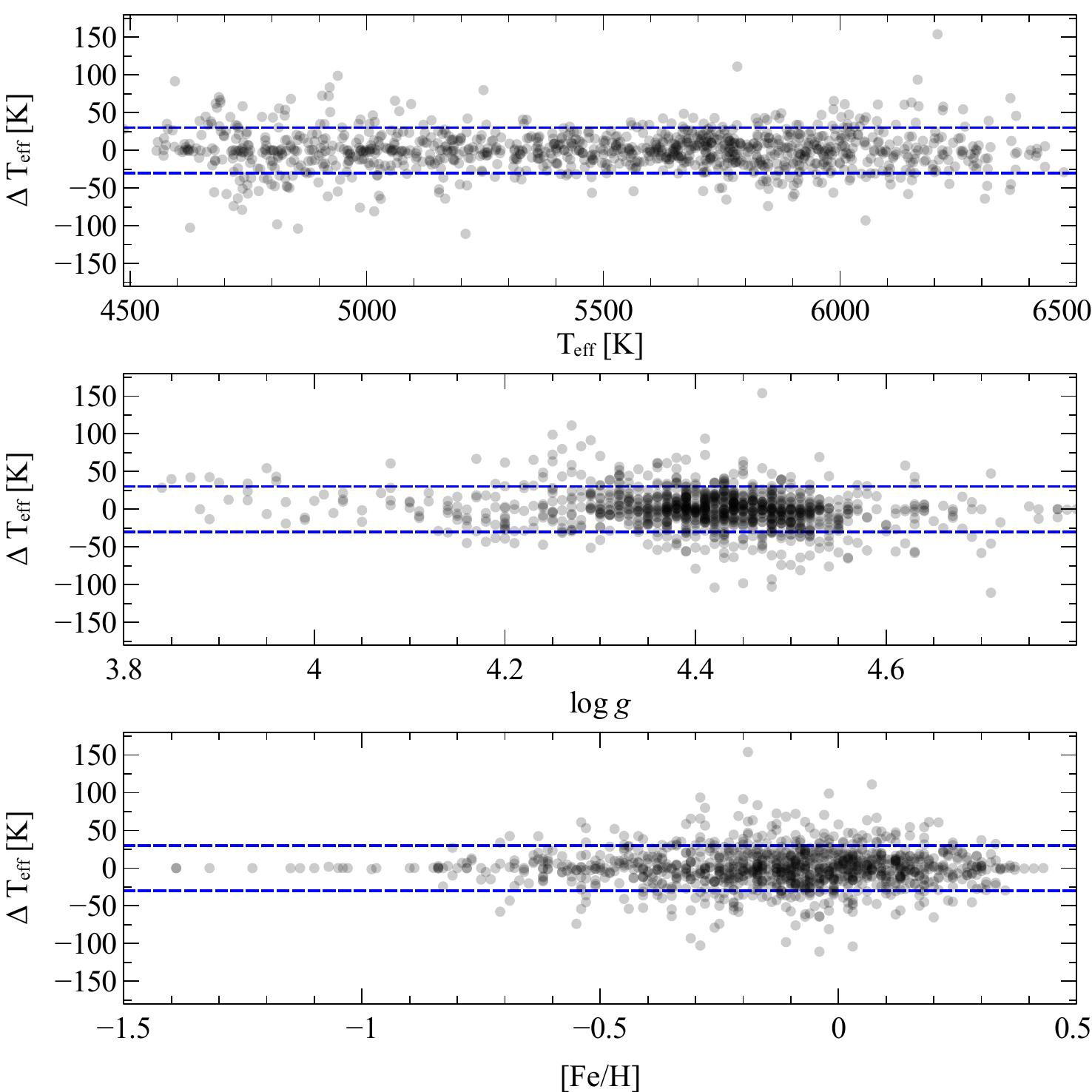}
\caption{The internal precision of temperature determination using our technique is shown by taking the difference between the values used  to determine \teff\ as a function of CCF areas, and the  values returned by the calibrated function Equation \eqref{eq:teff_func1} for the same CCF areas.  Blue lines show the precision of \EW measurements, \ie ,  the values used to calibrate the function, red lines show the $1-\sigma$  precision of our calibration.}
\label{fig:tg_intcal_dteff}
\end{figure}

\begin{figure}
  \includegraphics[width=\columnwidth]{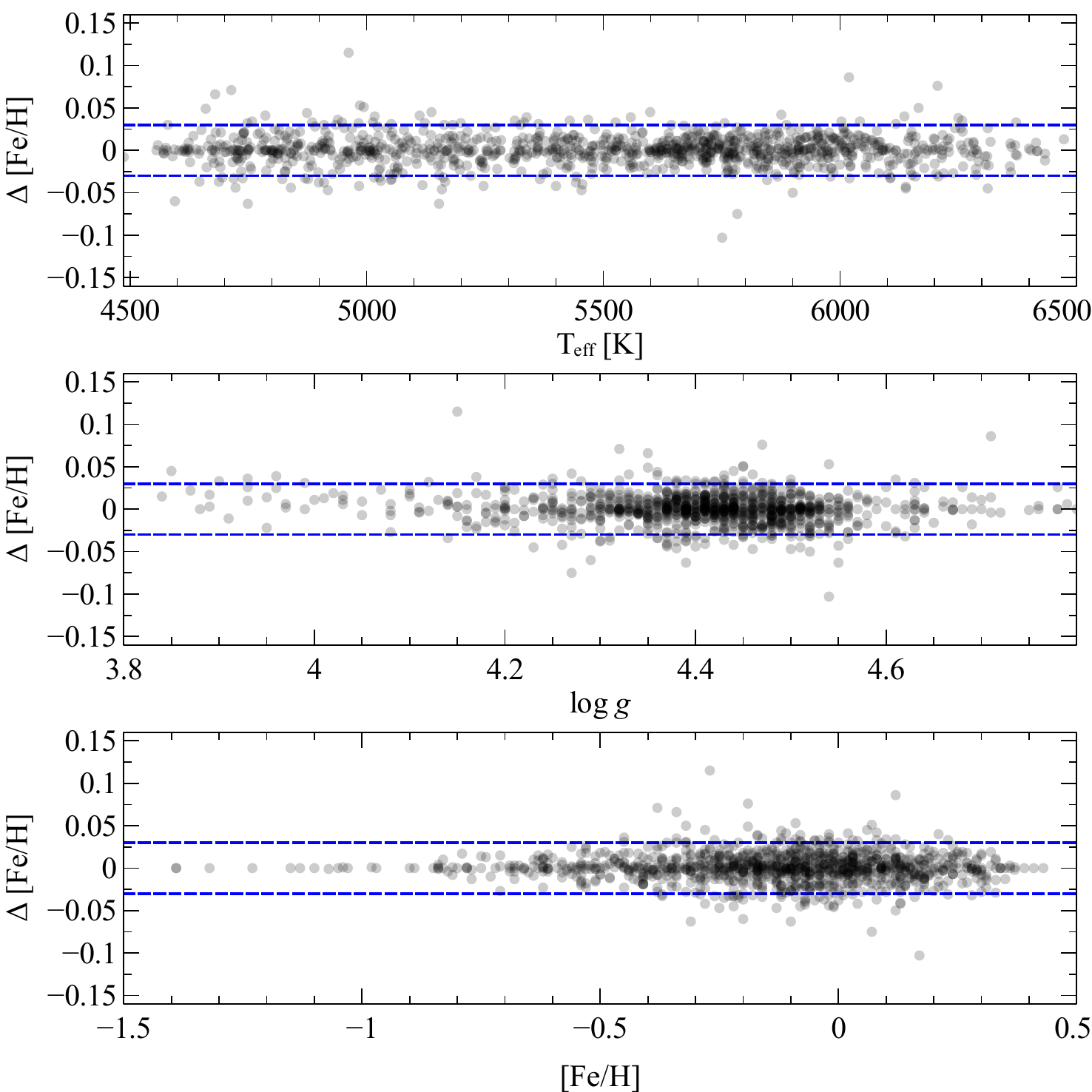}
\caption{As in Figure~\ref{fig:tg_intcal_dteff}, but for metallicity}
\label{fig:tg_intcal_dgfeh}
\end{figure}

The internal precision of the calibration is tested determining \teff\ and \gfeh\ from the same area values used to determine the function coefficients. Figure~\ref{fig:tg_intcal_dteff} and
Figure~\ref{fig:tg_intcal_dgfeh} show the difference between the stellar parameters used to calibrate Equations \eqref{eq:teff_func1} and \eqref{eq:gfeh_func1} as a function of the measured CCF areas, and the value returned by these functions when the same area values are given as input.

After exploring several possibilities, we decided to model $f_1$ and $f_2$ with three-dimensional Chebyshev polynomial functions of the first kind of order $6x6x6$. The Chebyshev polynomials are defined in the range $[-1,1]$, so the mid-points $A^{\textrm{mean}}$ and range $A^{\textrm{range}}$ for each variable must be chosen a priori. The variable transformation is defined by Equation \eqref{eq:area_to_areaprime}.

\begin{equation}\label{eq:area_to_areaprime}
A^{\prime} =( A^{\textrm{measured}} - A^{\textrm{mean}} )/ A^{\textrm{range}}
\end{equation}

This choice led to an internal precision comparable with the spectroscopic formal error of the atmospheric parameters, and the absence of any systematic trend in the residuals. We obtained $\sigma_{\textrm{\teff}} = 22$ K and $\sigma_{\textrm{\gfeh}} = 0.02~ \textrm{dex}$, against a reported precision of $\sigma_{\textrm{\teff}} = 30$ K and $\sigma_{\textrm{\gfeh}} = 0.03~ \textrm{dex}$ from \textit{EW} determination.

Once the coefficients of $f_1$ and $f_2$ have been determined, the only parameters required for \teff\ and \gfeh\ determination are:
\begin{itemize}
\item the three line-lists in Tables~\ref{tab:sousa2010_linelist};
\item the coefficients of the Functions \eqref{eq:teff_func1} and
  \eqref{eq:gfeh_func1};
\item the two parameters for each variables needed for Equation \eqref{eq:area_to_areaprime};
\end{itemize}
The CCFs must be computed with the same technique as described in Section~\ref{sec:CCF_mask}; since the areas are normalized to the continuum level, the size of the bin in the mask is not relevant. Flux correction plays a major role, since it is changing the  weights of single lines when assembling the CCF, and it should be performed as described in Section~\ref{sec:refraction-corr}.
Note that the general trend of stellar flux correction shown in Figure~\ref{fig:fluxcorr_star} is due to the technical characteristics of the telescope and spectrograph, and can be determined a priori. However, as a consequence the derived photospheric parameters will be affected by a larger uncertainties for not correcting night-by-night deviations from the general flux shape.

Generally speaking, the CCFs of the calibration sample and the CCF of a star with unknown parameters must  be computed following the same algorithm, to ensure a correct determination of the stellar parameters.

\section{Calibration of gravity}\label{sec:deriv-temp-surf}

In principle, all the lines are sensitive to pressure changes: with increasing density, a larger number of absorbers per volume is available and the atomic lines are stronger. For cool stars however the strengths of neutral lines of most elements do not depend much on gravity, as discussed by \cite{Gray:2005bb} (see again Figures~\ref{fig:tg_intcal_dteff} and Figure~\ref{fig:tg_intcal_dgfeh}).

On the other hand, lines arising from the ionized species of most elements are often very dependent on gravity. The rate of ionization of atoms depends on temperature, but the recombination rate is a function of temperature and gravity (Saha equation): recombination reduces the number of ionized atoms and it is faster with increasing pressure (gravity), thus lines from the most ionized states are very sensitive to this parameters.

In classical \textit{EW} analyses, an initial guess at gravity is made to determine \teff\ and \gfeh\ from neutral iron lines. Using the new determinations of the two parameters, the gravity estimate is improved by imposing ionization equilibrium of iron, \ie , the abundance derived from FeII lines must match that from FeI lines. The two steps are repeated until the three parameters converge.

We can proceed in a similar fashion to calibrate \logg\ as a function of CCF area. Following the approach in Section~\ref{sec:line_selection_CCF_mask} and Section~\ref{sec:model-temp-gfeh}, we define a CCF mask including isolated, unsaturated ionized lines for several elements, taken from \cite{Sousa:2010gl} and \cite{Neves:2009jz} and listed in Table~\ref{tab:sousa2010_linelist} (identified with the \texttt{mask} flag \texttt{I}).

\begin{figure}
 \includegraphics[width=\columnwidth]{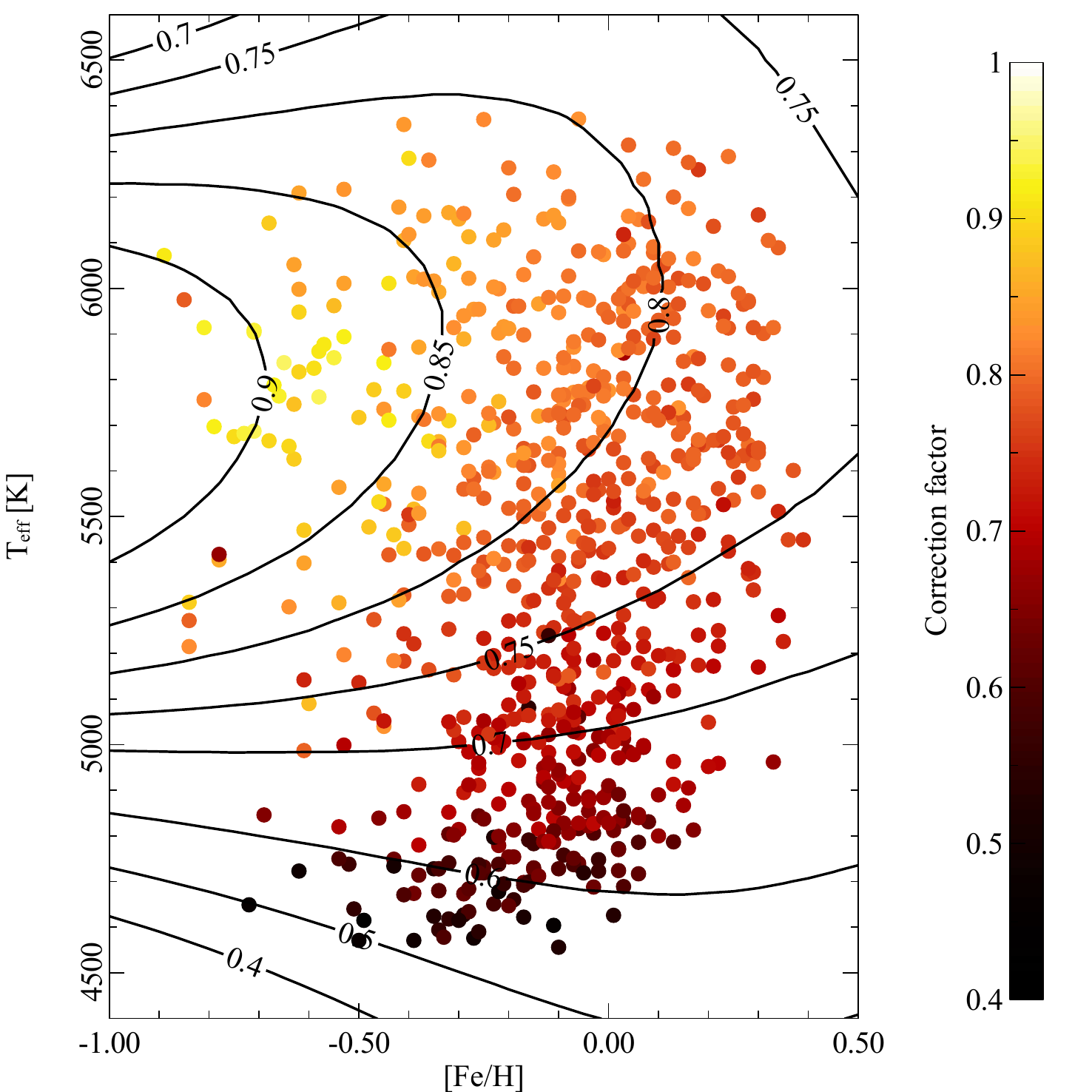}
\caption{The correction factor $F_{C}(\textrm{\teff},\textrm{\gfeh})$ required to rescale the synthesis-derived CCF areas to the actual values measured in stars. Only stars with $4.3 < \logg < 4.5$ have been included here. The resulting 2D Chebyshev polynomial fit is represented by the solid lines.}
\label{fig:FCcal_norm}
\end{figure}

We firstly attempted to calibrate gravity as a function of the three CCF areas from neutral lines plus the CCF area using ionized lines, as done with \teff\ and \gfeh, but the coverage in gravity of our sample was not sufficient to derive a direct calibration of \logg , not even when  \teff\ and  \gfeh\ were externally provided. Keeping in mind that the CCF area from the ionized lines mask is also a function of temperature and metallicity, we followed this strategy to calibrate \logg\ as a function of $A_{CCF}^{\star }$:

\begin{enumerate}
\item For each star, the CCF area using the ionized line mask  $A_{CCF}^{\star }$ is measured.
\item For each star, the expected CCF area from \EW s  $A_{\EW}^{\star }$  is determined using Equation \eqref{eq:ccf2ew_final} and the atmospheric stellar parameters from literature;
\item The correction factor $F_C = A_{\textrm{\CCF}}^{\star }/A_{\textrm{\EW}}^{\star }$ is determined using stars with $\textrm{\logg} = 4.4 \pm 0.1 $ and modeled with a 2D Chebyshev
polynomial function of the first kind (order $2 \times 2$) as a function of temperature and metallicity (Figure~\ref{fig:FCcal_norm}).
\item For each star, $A_{\textrm{\EW} }^{\textrm{\logg} }$ is evaluated for a grid of \logg\, values in the range $[3.6,4.9]$ and step 0.1 dex, as in step (ii).
\item $A_{\textrm{\EW} }^{\textrm{\logg} }$ are rescaled to the expected  $A_{CCF} ^{\textrm{\logg} }$  values by using the function $F_c$ determined in step (iii).
\item For each \logg\, grid point,  $A_{CCF} ^{\textrm{\logg} }$ is modeled  as a function of \teff\, and \gfeh\, with a 2D Chebyshev polynomial (order $3 \times 3$).
\end{enumerate}

The difference between synthetic and observed values for the CCF areas, modeled with the function $F_c$, is mainly due to the influence of nearby spectral lines which lower the observed CCF continuum but are not taken in account in Equation \eqref{eq:ccf2ew_final}. This explains why the ratio $F_C$ is systematically lower than unity and gets lower at cooler temperatures, where spectral lines grow in depth.

\EW s are calculated using the \texttt{ewfind} driver of \texttt{MOOG}. For the continuum determination, we decided to use the synthetic flux-calibrated stellar spectrum closest to the Sun from the \cite{Coelho:2005ab} stellar library, which has $\textrm{\teff}=5750$ K, $\textrm{\logg}=4.5~ \textrm{dex}$, $\textrm{\gfeh}=0.0~ \textrm{dex}$. The systematic flux variations introduced by using a single synthesis as reference for a broad range of spectral types is taken automatically into account in the calibration of $F_C(\textrm{\teff},\textrm{\gfeh})$.

Following the approach described above, to determine the stellar gravity \teff\, and \gfeh\, must be provided as input, using for example the values derived  in Section~\ref{sec:model-temp-gfeh}).
The values of $A_{\textrm{\CCF} }^{\textrm{\logg} }$ computed at \teff\, and \gfeh\, are then retrieved for each point of the \logg\, grid and interpolated, with \logg\ as a function of $A_{\textrm{\CCF} }$. Finally the gravity value is determined by using the $A^\star_{\CCF}$ measured from the spectrum with the ionized lines mask.

\begin{figure}
 \includegraphics[width=\columnwidth]{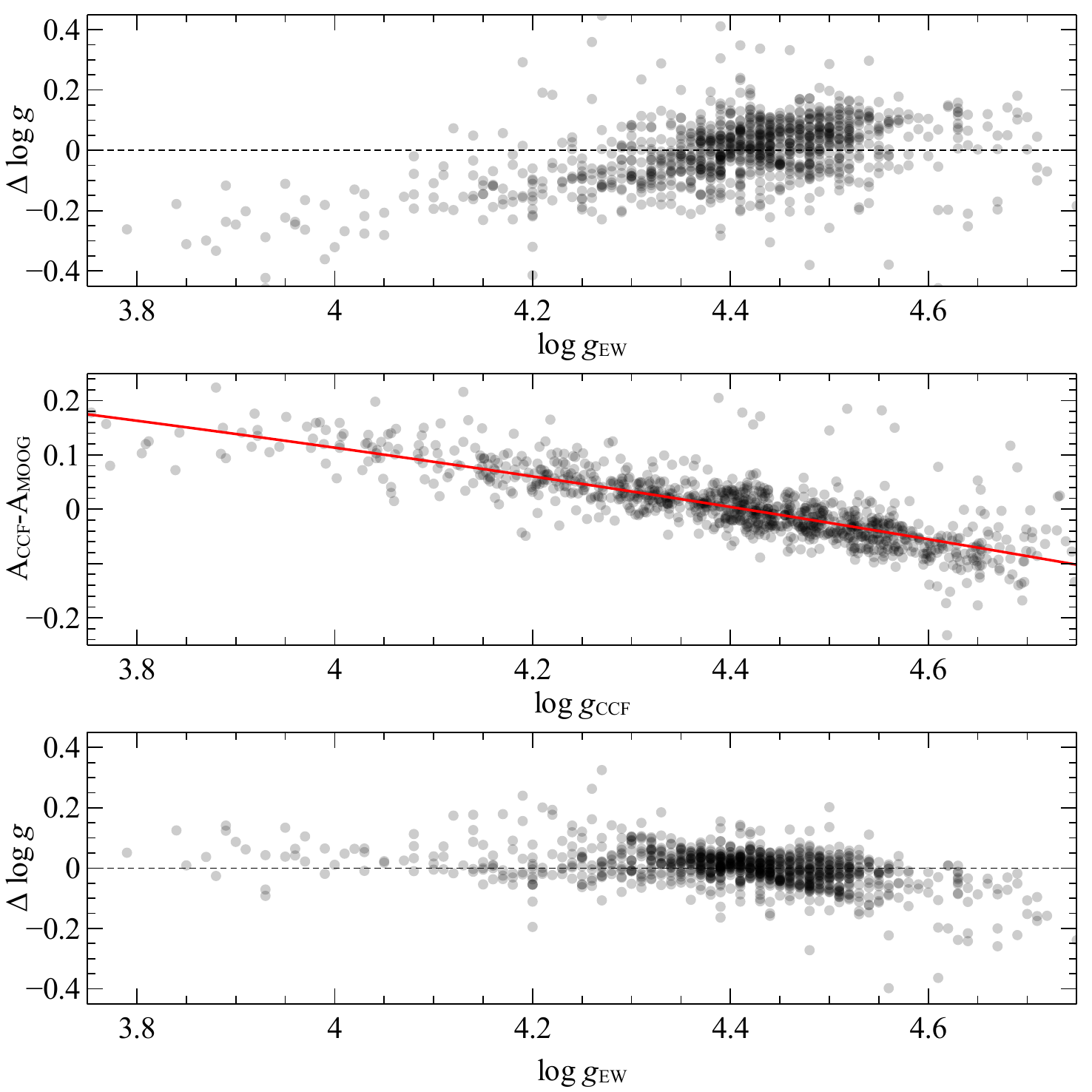}
\caption{Upper panel: The difference between CCF-derived and \EW -derived gravity is
plotted against gravity from literature.
Middle panel: the difference between the CCF area determined using the ionized lines mask and the expected synthetic value is plotted as a function of the gravity derived with the CCF approach. A quadratic fit is performed to model the systematic trend. Lower panel: The difference between the value
obtained from the CCF calibration after correcting for the systematic trend and the one from \EW\, analysis is plotted against literature gravity.}
\label{fig:logcal_trends}
\end{figure}

As done for \teff\, and \gfeh\, in Section~\ref{sec:model-temp-gfeh}, we test the internal precision of the calibration by comparing the derived \logg's  for the calibrators and the \logg $_{\textrm{\EW}}$ values from the literature. In the top panel of Figure~\ref{fig:logcal_trends}, the difference between these two values $\Delta \textrm{\logg}$  is plotted against the gravity from literature. A clear trend with the gravity of the star is present, possibly caused by our approach in determining the correction factor $F_{C}(\textrm{\teff},\textrm{\gfeh})$, which appears to be function of \logg\, as well. Since we do not have enough data to calibrate the correction factor as a function of gravity, a different approach has to be taken.

In the middle panel of Figure~\ref{fig:logcal_trends} the difference between the CCF area determined using the ionized lines  mask and the expected synthetic value is plotted as a function of the gravity derived from the CCF area. The red line represents a quadratic fit to the overall trend.
The scatter around this line is $\sigma{\Delta A}=0.026$, which is around $1-2\%$ of the average value of the CCF area. We decided to model such difference as a function of the derived gravity instead of the literature gravity so that we can apply this correction even for star without any a priori knowledge of the gravity.
Following this approach, an approximation of gravity is firstly determined using the measured CCF area, the area correction value is determined using the derived value of gravity and
added to the previously measured CCF area, and then this corrected value is used to get the final
estimate of gravity. The bottom panel of Figure~\ref{fig:logcal_trends} is a replica of the upper panel, with the CCF-derived gravity now determined after applying the model introduced in the middle panel.

\begin{figure}
 \includegraphics[width=\columnwidth]{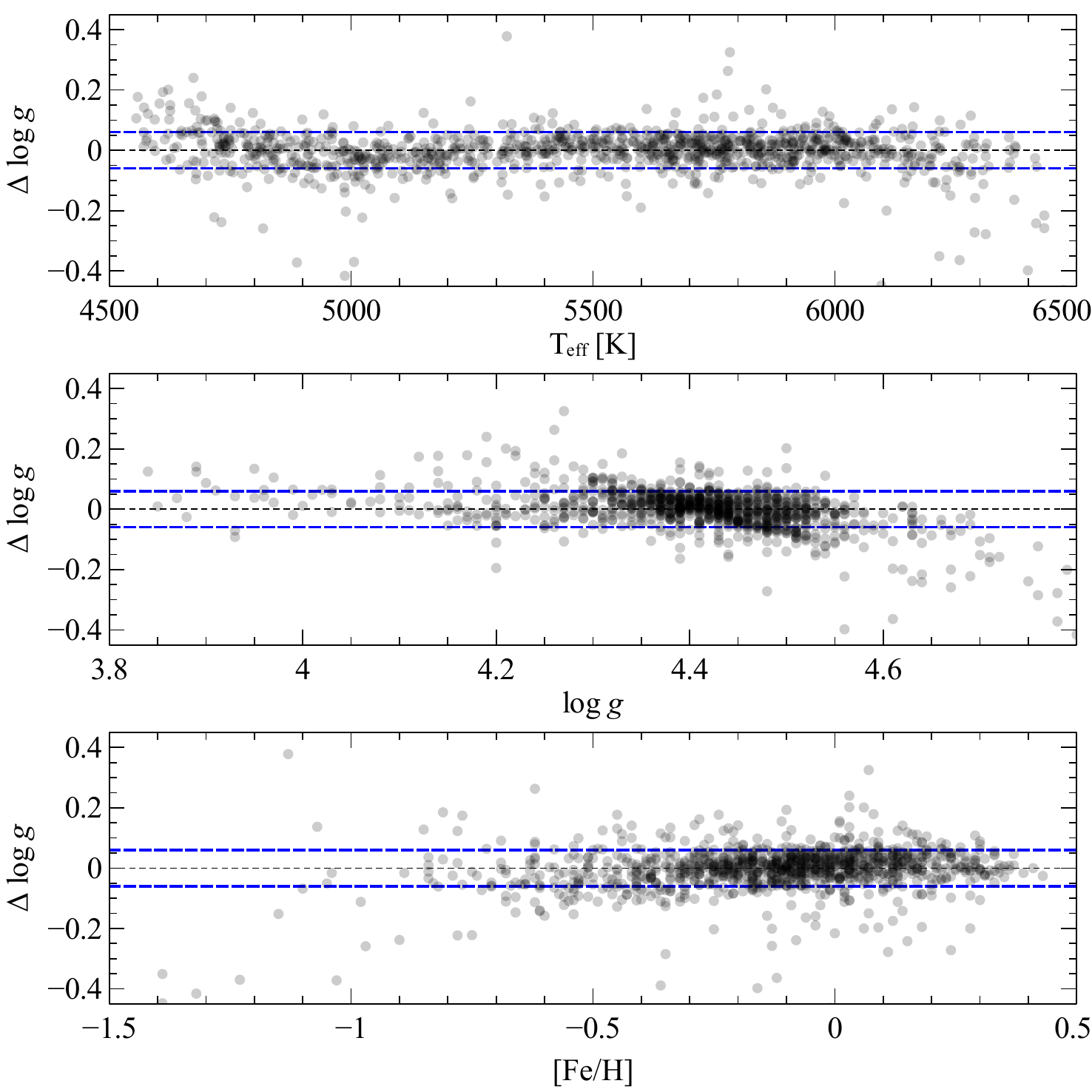}
\caption{As in Figure~\ref{fig:tg_intcal_dteff}, but for gravity}
\label{fig:logcal}
\end{figure}

Results for calibrators stars are presented in Figure~\ref{fig:logcal} in a similar fashion as done for temperature and metallicity. The dispersion in the residual distribution is of the same order of the average formal error ($\sigma_{\logg}=0.076 \,\textrm{dex}$ versus $<\sigma >_{\logg}=0.06 \,\textrm{dex}$), although a residual trend with temperature and metallicity is still present and at higher values of gravity and lower temperatures the calibration seems to be less reliable. While a more complex approach than the one presented in this section should be followed to correct for the systematic deviations in Figure \ref{fig:logcal_trends}, we note that the amplitude of the residual trend is smaller than the average error of the calibration stars (blue dashed lines in the Figure), meaning that no real improvement of the results would be obtained. Therefore we preferred to rely on our simpler but more robust approach.

\section{Atmospheric parameters as a function of \snr }\label{sec:glob-deriv-atmosph}

We applied the technique described in the previous sections to each individual exposure obtained with HARPS within the planet-search survey described in Section~\ref{sec:sample_stars}, and compared the outcome with the \EW-based atmospheric parameters obtained after stacking together the exposures. These observations have been gathered in different weather condition and sometimes with different integration times, but usually for a given star the observations are clustered around the \snr\, expected for average observing conditions. The number of observations varies largely from star to star, from a few observations to several hundreds, depending on the specific goal for that target (\eg , characterize its activity rather than discovering new planets). We analyzed a total of $\simeq 56000$ spectra, but we decided to retain for each star a maximum of 10 randomly selected determinations of the atmospheric parameters to better represent the dispersion of the parameters as a function of \snr , \ie , to avoid biasing the plot towards stars with many observations.

Temperature and metallicity are derived using the \textit{direct} calibration introduced in Section~\ref{sec:model-temp-gfeh}; gravity is then derived as described in Section~\ref{sec:deriv-temp-surf}, using the obtained stellar parameters as input. It is important to notice that these calibrations, although very precise, are reliable only in the range of parameters used to derive the functions \eqref{eq:teff_func1} and \eqref{eq:gfeh_func1}.

\begin{figure}
 \includegraphics[width=\columnwidth]{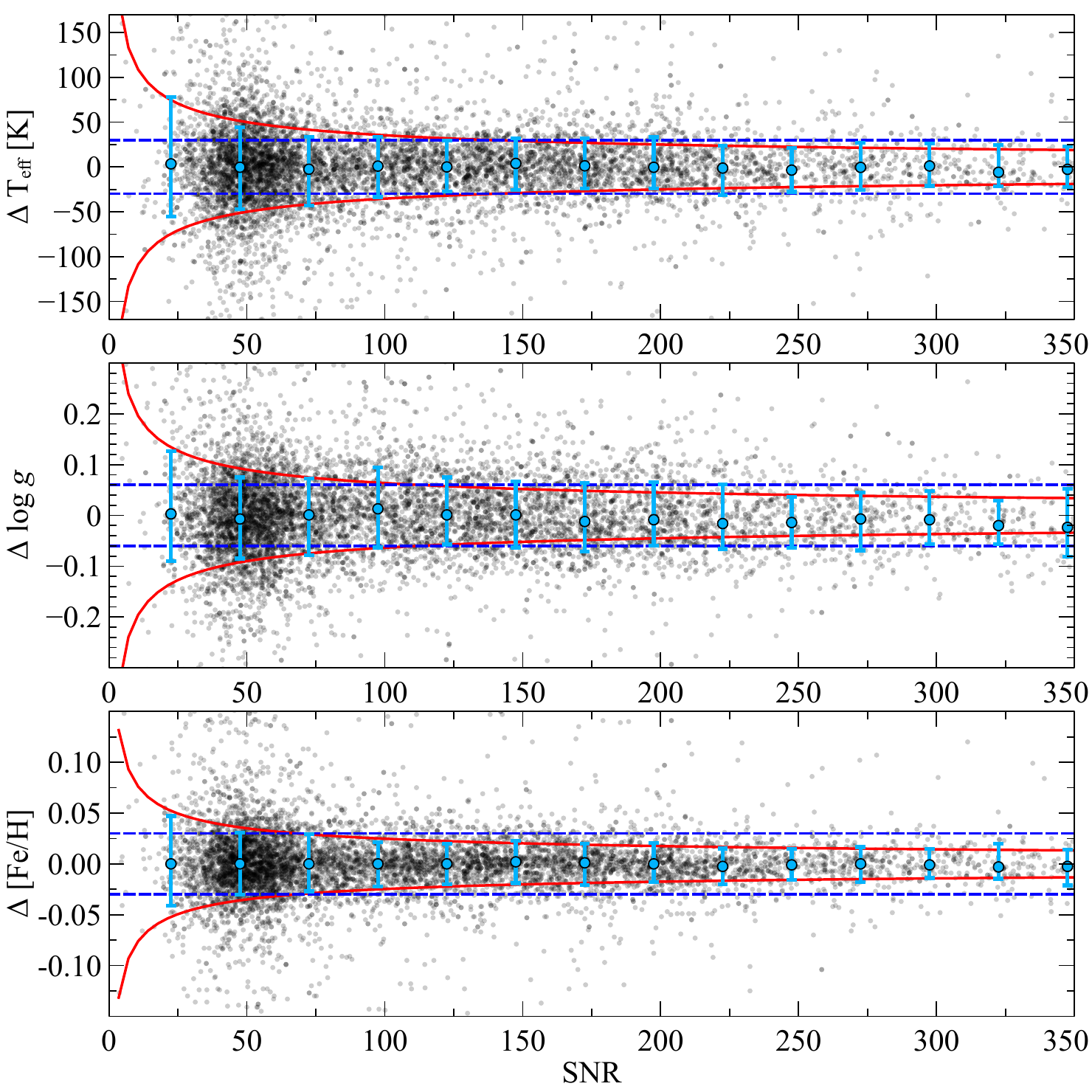}
 \caption{Difference between the CCF-derived atmospheric parameters for individual HARPS spectra and the \EW -derived parameters from high-\snr\, stacked spectra, as a function of the \snr . Temperature and metallicity have been derived using the \textit{direct} calibration, while for gravity we used the former two parameters as input.
Cyan points represent the $1 \sigma$ dispersion for several bins of SNR. The red line is the inverse square root law fit of the dispersion as a function of the SNR. Blue dashed lines represent the average error from \EW -derived parameters.}
\label{fig:ewfind_resall}
\end{figure}

To determine the precision as a function of \snr, the $\sigma$ of the distribution of the difference has been computed for several bins of \snr. At low \snr\ a simple inverse square-root law of \snr\, provides a very good fit of the points, proving that our measurements are photon-limited. For $\textrm{\snr}=50 $ we have a precision of $\sigma_{\textrm{\teff}} = 50$ K, $\sigma_{\textrm{\logg}} = 0.09~ \textrm{dex}$ and $\sigma_{\textrm{\gfeh}} =0.035~ \textrm{dex}$, while the accuracy of the technique is provided
by the accuracy of the atmosphere parameters of the calibrators. The precision of our technique is limited by the goodness of the calibration at a \snr\, between 100 and 150, so for \snr\, similar or greater than these values the average error from \EW\, analysis should be considered as an estimate for the associated errors.  \\

\section{Discussion}\label{sec:discussion}

We have presented a new technique to quickly determine reliable stellar atmosphere parameters for FGK Main Sequence stars using several CCFs specifically built to be more sensitive to a given atmosphere parameter respect to the others. Our approach relies on a set of high-resolution, high \snr\, stars already classified spectroscopically, which defines the limit on the accuracy of our technique.

We have developed this approach with two goals in mind.
The first goal is to enable the spectroscopic classification of stellar objects at observing time, a few seconds after the end of the exposure, with a precise tool that is quick and easy to use at the same time.  This tool can be extremely useful when performing time series of poorly characterized target, \eg , follow-up of faint planet-host candidates, to exclude stars that do not match a criteria selection  (\eg , stars that are not solar-type stars) after one or two observations with  a considerable optimization of telescope time.
The second goal is to determine atmospheric parameters of faint objects observed at very low \snr\, with high-resolution spectroscopy, when an equivalent width analysis is still not feasible even after co-addition of all the available spectra.

To find the relationship between CCFs parameters and photosphere parameters we have used a dataset of high-resolution, high \snr\, spectra of 1111 stars with accurate parameters derived using equivalent widths from literature. The spectrum of each star is actually the result of a co-addition of several spectra at lower \snr, since these data have been gathered for exoplanet search. The lower \snr\, spectra have allowed us to understand the effective precision of our technique on real data.

Two different calibrations have been presented. The first is purely empirical and it provides temperature and metallicity from the observed CCF areas for stars with parameters in the ranges $\textrm{\teff} \simeq [4500,6500]$ K, $\textrm{\logg} \simeq [4.2,4.8]$ and  $\textrm{\gfeh} \simeq [-1.0,0.5]$  (solar-type and slightly evolved stars). We achieve a precision of
$\sigma_{\textrm{\teff}} =50$ K and $\sigma_{\textrm{\gfeh}} = 0.035~ \textrm{dex}$ at $\textrm{\snr}=50$, with the precision being an inverse square-root law of \snr .
The second calibration is based on the transformation of synthetic equivalent width in CCF areas and it allows the determination of gravity when temperature and metallicity are provided as input. For gravity we reach a precision of  $\sigma_{\textrm{\logg}} = 0.09~ \textrm{dex}$, with this value taking into account the errors in \teff\ and \gfeh\ introduced by the empirical calibration. In both cases the precision for HARPS spectra with \snr\ $\gtrsim$ 100 and the overall accuracy are limited by the set of stars used as reference. Given the very high \snr\ of the calibration sample, better performance can be achieved by expanding the parameter space covered by the calibration stars, \eg , by including stars at lower metallicity, rather than increasing the number of observations of the stars already in the sample.

We have developed this open-source tool\footnote{Available at \url{https://github.com/LucaMalavolta/CCFpams}} for HARPS and HARPS-N data and for FGK Main Sequence  stars, but our approach can be easily extended to other instruments with similar or larger spectral range and similar resolution, or to other spectral range and stars with different characteristics (\eg, Red Giant Branch stars or M dwarfs) if a large sample of reference stars is available to calibrate the CCFs as a function of photospheric parameters. When either of the two cases above does not apply, we provide the mathematical formulation required to transform synthetic \EW s to CCF areas. Our tool will allow an easy, quick and reliable characterization of candidate transiting planets from present and  future transit surveys such as NGTS \citep{Chazelas2012}, TESS \citep{Ricker2014} and PLATO \citep{Rauer2014}.

\section*{Acknowledgments}
The research leading to these results received funding from the European Union Seventh Framework Programme (FP7/2007- 2013) under grant agreement number 313014 (ETAEARTH).
LM is grateful to the Observatory of Geneva for its financial support as visiting researcher.
Partial support for this work has been provided by the US National Science
Foundation under grants AS1211585 and AST-1616040.

\bibliographystyle{mn2e}
\bibliography{Malavolta_CCF_resubmitted_plain}
\bsp
\label{lastpage}
\end{document}